\documentclass[twocolumn,showpacs,preprintnumbers,amsmath,amssymb]{revtex4}

\usepackage{graphicx}
\usepackage{dcolumn}
\usepackage{bm}

\begin{document}

    \title{Metamorphosis of the Quantum Hall Ferromagnet at $\nu = 2/5$}
\author{Michelle Chen$^{1}$, Woowon Kang$^{1}$ and Werner Wegscheider$^{2}$}
\affiliation{$^{1}$James Franck Institute and Department of 
Physics, University of Chicago, Chicago, Illinois 60637\\
$^{2}$Institut f\"{u}r Experimentelle and Angewandte Physik, 
Universit\"{a}t Regensburg, D-93040 Regensburg, Germany and 
Walter Schottky Institut, Technische Universit\"{a}t 
M\"{u}nchen, Am Coulombwall, D-85748 Garching, Germany}

\date{\today}

\begin{abstract}
We report on the dramatic evolution of the quantum Hall ferromagnet in the 
fractional quantum Hall regime at $\nu = 2/5$ filling. A large enhancement in 
the characteristic 
timescale gives rise to a dynamical transition into a novel quantized Hall 
state. The observed Hall state is determined 
to be a zero-temperature phase distinct from the spin-polarized and spin-unpolarized 
$\nu = 2/5$ fractional quantum Hall states. It is characterized by a strong 
temperature dependence and puzzling correlation between temperature and time.
\end{abstract}


\maketitle

In two-dimensional electron systems under strong transverse magnetic 
field, transitions to quantum Hall ferromagnets occur when two degenerate 
quantum Hall states coincide within some configurational 
space\cite{GirvinMacDonald,Jungwirth98,Jungwirth01A}. The correlation between 
the two degenerate quantum Hall phases can be treated theoretically - in analogy 
to isospin degeneracy in nuclear physics - through a pseudospin spinor 
representation of the relevant quantum numbers\cite{GirvinMacDonald}.
Easy-axis quantum Hall ferromagnets have been shown to emerge in systems 
in which two degenerate quantum Hall states that differ primarily in the 
spin part of the wave function\cite{Jungwirth98,Jungwirth01A}. 
Near a crossing between two spin levels, competition between two 
degenerate quantum Hall states produces a correlated ground state 
termed a quantum Hall ferromagnet (QHF).
Experimental studies in a number of systems have shown that Ising 
quantum Hall ferromagnets are characterized by  metastability and 
hysteresis\cite{Jungwirth98,Cho98,Kronmuller98,Kronmuller99,
Piazza99,Eom00,DePoortere00,Smet01}. 

Of the various examples of easy-axis QHFs, 
the $\nu = 2/5$ fractional quantum Hall effect (FQHE) state in the 
crossover region between the spin-unpolarized and spin-polarized 
states demonstrates the transport signatures associated 
with the Ising quantum Hall ferromagnetism. Realized 
under the conditions of reduced $g$-factor 
arising from application of hydrostatic pressure in excess of 10 kbar, 
the transport in proximity of $\nu = 2/5$ filling  
exhibits strong hysteresis\cite{Cho98} and anomalous logarithmic 
time-dependence\cite{Eom00}.
The path-dependent order parameter of an Ising QHF 
provides an elegant explanation for the hysteretic 
behavior\cite{Jungwirth98,Cho98}. Local frustration of the 
pseudospin domains gives rise to a glassy behavior and the associated 
logarithmic time-dependent transport\cite{Eom00}. 
Competition between the local
interaction energy of the ferromagnetic domains and the configurational 
entropy of the domains walls is thought to be responsible for the 
large enhancement in the characteristic timescale.
An important unresolved question concerning the glassy dynamics of the 
QHF at $\nu = 2/5$ involves the nature of its ground state in the 
$t \rightarrow \infty$ limit.

In this paper, we report on a novel phase transition of the QHF
at $\nu = 2/5$ in the crossover regime between the 
spin eigenstates of the $\nu = 2/5$ FQHE.
After an extensive duration of annealing, the time-dependent
aging features previously attributed as the relaxation of QHF domains
culminates with a quantized  Hall state at $\rho_{xy} = 5h/2e^{2}$. 
Due to large enhancement in the 
characteristic time associated with the QHF at 
$\nu = 2/5$, we are able to capture the progressive evolution
into a highly correlated quantum Hall phase.
This phase differs from the spin-singlet and spin-polarized 
$\nu = 2/5$ FQHE states in that  
(1) initially there is no quantum Hall state 
at $\nu = 2/5$ under $\sim$13.5 kbar of pressure, (2) the quantized
Hall state only appears after $\sim$2 days of annealing, (3) it 
exhibits hyper-sensitivity 
to changes in temperature with an increase of  $\sim 10$mK sufficient 
to destabilize the quantum Hall state, and (4) it demonstrates 
an anomalous correlation of magnetoresistance as a function of 
temperature and time. Based on these findings, we conclude that the 
quantized Hall state obtained via annealing is a zero temperature 
correlated phase distinct from the spin-singlet and the spin-polarized FQHE 
states at $\nu = 2/5$.

High-quality GaAs/AlGaAs heterostructure was placed under large 
hydrostatic pressure inside a miniature copper-beryllium pressure 
clamp. The detail of the pressure apparatus has been described 
elsewhere\cite{Cho98,Eom00}. 
Application of large hydrostatic pressure reduces the magnitude of the 
Lande $g$-factor and the consequent reduction in the Zeeman energy, 
$E_{Zeeman} = g\mu_{B}B_{total}$, enhances the spin-reversals of 
electrons.
In order to optimize the competition between the spin-polarized and 
spin-unpolarized $\nu = 2/5$ FQHE states, the pressure was tuned to 
the crossover pressure of $\sim$13.5 kbar.
The density of the sample under 13.5 kbar of pressure is 
$n = 1.2 \times 10^{11}cm^{-2}$ with 
mobility of $\mu = 1.8\times10^{6}cm^{2}/Vs$.
Typical annealing run lasted $\sim$2 days prior to 
subsequent magnetic field sweeps. The pressure inside the 
pressure-cell remained stable at low temperatures and no change could be detected 
even after several months of experiment. All 4 samples studied during the 
course of the experiment exhibited the reported effect.

\begin{figure}
\includegraphics[width=86mm]{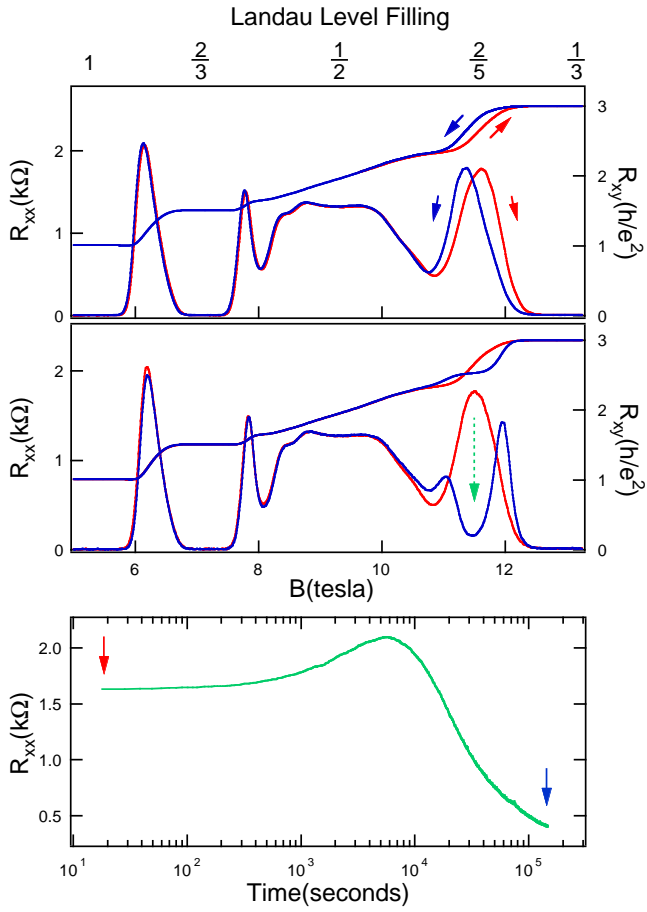}
\caption{History dependent transport of the quantum 
Hall ferromagnet at $\nu = 2/5$. The data was obtained under 13.5 kbar 
of pressure.
(a) Hysteresis in magnetotransport associated with the quantum Hall 
ferromagnet at $\nu = 2/5$. 
(b) Comparison of the magnetotransport before (red) and after 
(blue) annealing of $1.5 \times 10^{5}$ seconds.
Only the down field sweeps are shown for clarity.
(c) Time dependence of magnetoresistance at $\nu = 2/5$ from annealing 
at temperature of 30mK.
\label{fig:rxxevo1}}
\end{figure}

Fig. \ref{fig:rxxevo1}a illustrates the representative
hysteresis in magnetoresistance and Hall 
effect that appears at $\nu = 2/5$ filling under 13.5 kbar of pressure. 
Although no quantum Hall state at 
$\nu = 2/5$ is found, there is a prominent hysteresis as both 
magnetoresistance and Hall effect depend on the direction and sweep 
rate of the applied magnetic field. 
Fig. \ref{fig:rxxevo1}b compares the  magnetoresistance and Hall effect 
of a specimen before and after approximately $\sim$2 days annealing  at 
temperature of 30mK.
Fig. \ref{fig:rxxevo1}c illustrates the time dependence of the 
magnetoresistivity at 
$\nu = 2/5$ under temperature of 30mK and 11.6 tesla of magnetic field. 
The resistance initially increases until it reaches a maximum
around $\sim 1\times 10^{4}$ seconds.  Subsequent resistance decay
is slow but persistent even after 2 days of annealing. Magnetic field 
sweeps performed after annealing reveal an unexpected quantum Hall state 
as evidenced from the pronounced minimum in the magnetoresistance and 
the quantized Hall plateau with $\rho_{xy} = 5h/2e^{2}$. This shows 
that the preferential ground state after an extensive period 
annealing of the QHF at $\nu = 2/5$ is a quantized Hall state.

\begin{figure}
\includegraphics[width=86mm]{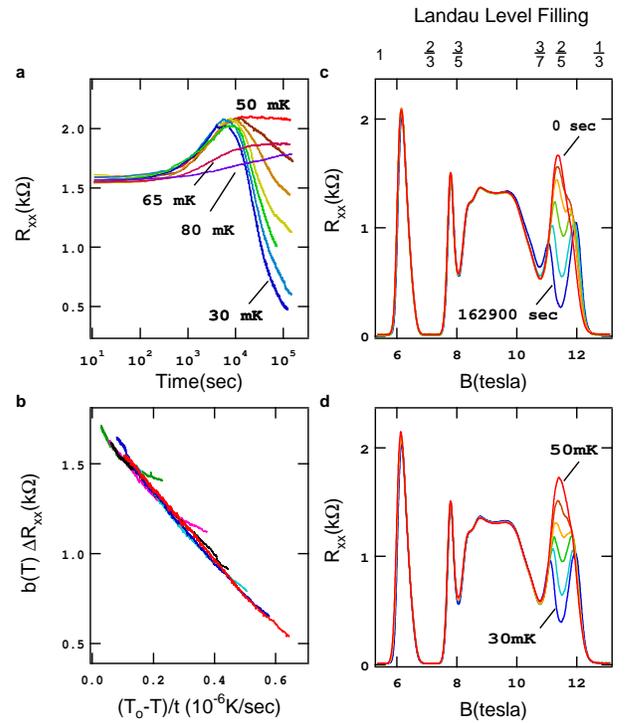}
\caption{Magnetoresistance at $\nu$ = 2/5 as a function of
temperature and time.
(a) Time dependence of magnetoresistance for annealing at different 
temperatures. From top to bottom, the traces correspond to 50 (red), 
47, 43, 40, 35, 34, and 30 mK (blue) from top to bottom.
(b) Scaling analysis of the change in magnetoresistance above 
$2\times 10^{4}$ seconds from (a).
The change in magnetoresistance, $\Delta R_{xx}$, for different 
temperatures can be collapsed onto a universal curve of the form 
$b(T) \Delta R_{xx} \sim (T_{\circ} - T)/t$. $b(T)$ is the 
the size of $\Delta R_{xx}$ and $T_{\circ}$ is the characteristic 
temperature of $\sim 50 mK$. 
(c) Magnetoresistance in the lowest Landau level obtained after 
different duration of annealing. 
=From top to bottom, the annealing duration corresponds to 
0 (red), 36700, 65700, 94300, 132700, and 162900 seconds (blue).
(d) Magnetoresistance obtained after $\sim 1.5\times 10^{5}$ seconds 
for temperatures of 50 (red), 47, 43, 40, 34, 30 mK (blue). 
\label{fig:rxxevo2}}
\end{figure}

Fig. \ref{fig:rxxevo2}a illustrates the time dependence of the 
magnetoresistance obtained through annealing at $\nu = 2/5$.
The time dependent magnetoresistance for different temperatures
can be separated into two distinct behaviors before and after the 
resistance peak found near $t_{\circ} \approx  1\times 10^{4}$ seconds. 
In the initial part of annealing during $t < t_{\circ}$, magnetoresistance 
is weakly temperature dependent and increases slightly between 30 to 
50mK.
For $t > t_{\circ}$, the $R_{xx}$ at $\nu = 2/5$ exhibits a strong 
temperature dependence associated with the fromation of the quantized Hall 
state at $\nu = 2/5$. A resistance decrease of $\sim$75\% by the end of 
the annealing run at $1.5\times 10^{5}$ seconds
is found at 30mK. Raising the
temperature beyond 50mK removes the resistance downturn above $t > 
t_{\circ}$ and produces the logarithmic time dependence reported 
earlier\cite{Eom00}. Based on these findings, we attribute the 
logarithmic time dependence as a feature of high temperature annealing of 
the QHF at $\nu = 2/5$.

Fig. \ref{fig:rxxevo2}b shows the quantitative relationship between 
time and temperature in the annealed quantum Hall state at $\nu = 2/5$.
The change in magnetoresistance, $\Delta R_{xx}$, for $t > 10^{4}$ 
seconds from Fig. \ref{fig:rxxevo2}a can be collapsed onto a universal 
curve of the form 
$b(T) \Delta R_{xx} \sim (T_{\circ} - T)/t$, where $b(T)$ is the 
temperature dependent prefactor that is equivalent to
the size of $R_{xx}$ 
reduction. The characteristic temperature, $T_{\circ}$, of 
$\sim 50 mK$ provides a measure of the energy scale of the  
phase. This inverse correlation between time and temperature 
demonstrates a remarkable interplay of the dynamics and 
thermodynamics of the annealed, low temperature phase.

Fig. \ref{fig:rxxevo2}c provides a sequence of snapshots of the quantum 
Hall phase at $\nu = 2/5$ with increasing annealing time. The quantum Hall 
state becomes stronger as longer annealing is applied.
Emergence of the new Hall state is possible as long as the annealing 
is performed in proximity of $\nu = 2/5$. From the onset of $R_{xx}$ 
minimum at $\nu = 2/5$, we estimate the characteristic time of the 
transition to be $\sim 1 \times 10^{5}$ seconds, enormously enhanced 
compared to the single electron relaxation time of $\sim$$10^{-12}$ seconds. 
Fig. \ref{fig:rxxevo2}d shows the representative traces at different
temperatures attained after $\sim 1.5\times 10^{5}$ seconds of annealing. 
The annealed Hall state at $\nu = 2/5$ exhibits an uncharacteristically 
strong sensitivity to temperature compared to the neighboring FQHE 
states. Raising the temperature to $\sim$50mK is sufficient 
to displace the quantized Hall state at $\nu = 2/5$. Interestingly,
the temperature dependence of magnetoresistance is nearly a 
mirror-image of the time-dependence measurement of Fig. \ref{fig:rxxevo2}c
with temperature acting similar to inverse time. This presumably 
reflects the scaling relation between temperature and time
established in Fig. \ref{fig:rxxevo2}b.

The time dependence of $R_{xx}$ in Fig. \ref{fig:rxxevo2}a can be 
phenomenologically modeled in terms of two competing processes 
related to the motion of QHF domains, in analogy to the
slow dynamics found in systems such as structural glass\cite{Debenedetti96} 
and spin glass\cite{Fischer91}. Within the framework of the droplet 
scaling theory of spin glasses\cite{Fisher88}, domain growth in glassy
systems involves overcoming the barriers whose height grows with 
increasing lengthscale. 
The early, temperature-independent
part of annealing may be explained in terms of coarsening of QHF domains. 
The initial rise of resistivity may be due to novel one-dimensional excitations
responsible for the dissipation within the domains walls of
QHF\cite{Jungwirth01}. The latter stage of annealing appear to be 
dominated by the second, slower process that reduces dissipation and 
eventually leads to the quantized Hall phase. The $R_{xx}$ peak at 
$t = t_{\circ}$ may involve formation of a critical state in which a QHF
domain with some preferential pseudospin vector percolates throughout 
the sample. This would greatly accelerate the growth of the domains 
along with reduced dissipation as the total length of the domain 
wall is decreased. Appearance of the low temperature quantum Hall 
phase will continue until the domain size becomes comparable to 
the sample dimension. 

An unusual feature of this phase involves its extreme
sensitivity to temperature. While it may take a day or more of 
annealing to realize the quantum Hall state at $\nu = 2/5$, it is 
remarkable that raising the temperature by  
$\sim$10 mK is sufficient to 
destabilize the Hall phase. Such an uncharacteristically strong 
effect of temperature can be understood if the annealed QHF
is a zero temperature phase whose quantum mechanical 
coherence is readily destabilized by finite temperature. In terms of 
the QHF domains,
thermal excitations produce local reversal of pseudospin magnetization 
and nucleate domains that destroy the underlying long range coherence 
achieved through annealing. This apparent ease of destabilization at 
low temperature points to smallness of its gap associated with 
its unique correlated origin.

\begin{figure}
\includegraphics[width=86mm]{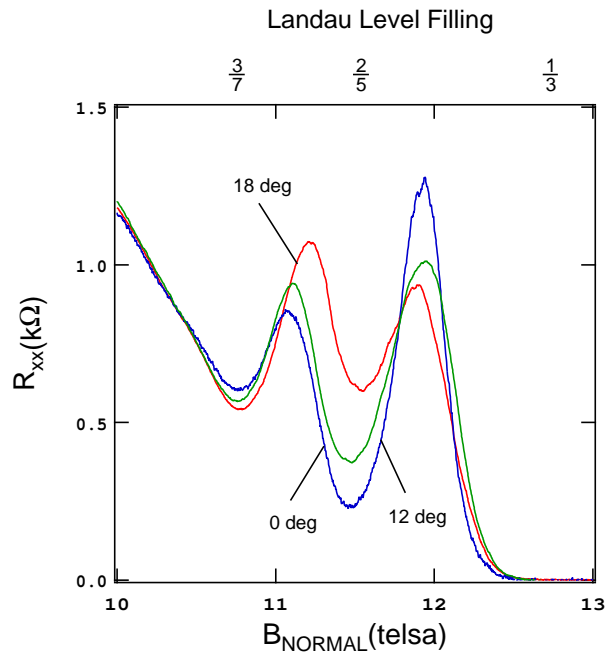}
\caption{Angular dependence of the $\nu$ = 2/5 fractional quantum Hall 
state at temperature of 40mK. The sample was annealed for $\sim$48 hours
prior to the field sweeps. The resistance minimum 
at $\nu = 2/5$ rises slightly with increasing in-plane magnetic field. 
\label{fig:rxxevo3}}
\end{figure}

In Fig. \ref{fig:rxxevo3}, we show the effect of in-plane magnetic 
field on the annealed phase at $\nu = 2/5$ at a temperature of 40mK.
With increasing tilt-angle, the annealed phase grows weaker as the 
$R_{xx}$ minimum for $18^{\circ}$ is about factor of 2 larger than the 
minimum for $0^{\circ}$. 
The weakening of the annealed quantum Hall phase at $\nu = 2/5$ 
can be interpreted as reduction in its energy gap under a larger 
Zeeman energy, suggesting that the spin-polarization of the 
annealed phase may be a spin-singlet. However, such an
interpretation is  complicated by the fact that presence of an
in-plane magnetic field creates an imbalance between the energies of 
the spin-polarized and spin-unpolarized states at $\nu = 2/5$ and this 
may produce a different annealing path at finite tilt angles 
compared to $0^{\circ}$. While the question regarding its spin 
polarization cannot be unambiguously answered, the tilted field study 
nonetheless shows that in-plane magnetic field tends to destabilize 
the annealed quantum Hall phase at $\nu = 2/5$.

In the context of spin transitions within the composite fermion model 
of the FQHE, it has been proposed that a new type of charge and spin 
density wave states can arise at $\nu = 2/5$ at intermediate values of 
Zeeman coupling. An admixture of two spin bands of composite fermion Landau 
levels at $\nu = 2/5$ can give rise to a partially polarized density 
wave (PPDW) state\cite{Murthy00}. Unlike the $\nu = 2/5$ FQHE state 
which can be either polarized or unpolarized, a PPDW possesses a 
spin polarization at half the value of the maximal polarization. 
Since PPDW states are stabilized under the conditions - of zero temperature 
and vanishing Zeeman energy - realized in our experiment, a possible 
consequence of annealing may involve a phase transition to a PPDW state.  
In addition, the angular dependence shown in Fig. \ref{fig:rxxevo3} 
is also consistent with presence of a PPDW. A larger Zeeman 
energy increases the spin-polarization of a PPDW from the half of maximum 
polarization, leading to its eventual destabilization at large tilt angles.
However, as discussed earlier, the effect of in-plane field on the annealing 
process needs to be properly accounted.

In summary, we have observed a transition into a novel quantum Hall
phase at $\nu = 2/5$ in the limit of small $g$-factor and
large time domain. Found at 
the level crossing between spin-polarized and spin-unpolarized $\nu = 2/5$ 
FQHE state, previously reported metastable state associated with 
the quantum Hall ferromagnetism evolves into a zero temperature 
correlated phase. This phase is distinguished by
large enhancement of the characteristic time, a bizarre correlation 
between temperature and time and its high sensitivity to thermal fluctuations.  
Within the coarsening picture of QHF domains, the observed quantum 
Hall state may represent a QHF state with long range spatial 
coherence.  Alternatively it may be the postulated partially polarized 
density wave or some other type of quantum Hall state.

We would like to thank A.H. MacDonald, G.F. Mazenko, and G. Murthy for 
useful discussions. The work at the University of Chicago is supported by
NSF DMR-9808595 and DMR-0203679.


\end{document}